\documentclass[aps,prb,reprint,superscriptaddress,showpacs]{revtex4-1}

\usepackage{graphicx}
\usepackage{amsmath}
\usepackage{color}
\usepackage{array}

\usepackage{ulem}

\begin{document}

\title{
Generalized correction to embedded-atom potentials for modeling equilibrium and non-equilibrium properties of metals
}

%%%%%%%%%%%%%%%%%%%%%%%%%%%%%%%%%%%%%%%%%%%%%%%%%%%%%%%%%%%%%%%%%%%%%

\author{Alexey Verkhovtsev}
\email[]{verkhovtsev@mbnexplorer.com}
\altaffiliation{On leave from Ioffe Institute, Politekhnicheskaya 26, 194021 St. Petersburg, Russia}
\affiliation{MBN Research Center, Altenh\"oferallee 3, 60438 Frankfurt am Main, Germany}
%\affiliation{German Cancer Research Center (DKFZ), Division of Medical Physics in Radiation Oncology, Im Neuenheimer Feld 280, 69120 Heidelberg, Germany}

\author{Andrei V. Korol}
\affiliation{MBN Research Center, Altenh\"oferallee 3, 60438 Frankfurt am Main, Germany}
\affiliation{Department of Physics, St. Petersburg State Maritime Technical University, Leninsky ave. 101, 198262 St. Petersburg, Russia}

\author{Gennady Sushko}
\affiliation{MBN Research Center, Altenh\"oferallee 3, 60438 Frankfurt am Main, Germany}

\author{
\framebox[1.0\width]{Stefan Schramm}}
%Stefan Schramm}
\affiliation{Frankfurt Institute for Advanced Studies, Goethe-Universit\"at, Ruth-Moufang-Str. 1, 60438 Frankfurt am Main, Germany}

\author{Andrey V. Solov'yov}
\altaffiliation{On leave from Ioffe Institute, Politekhnicheskaya 26, 194021 St. Petersburg, Russia}
\affiliation{MBN Research Center, Altenh\"oferallee 3, 60438 Frankfurt am Main, Germany}

\date{\today}

\begin{abstract}
A modification of an embedded-atom method (EAM)-type potential is proposed for a quantitative description
of equilibrium and non-equilibrium properties of metal systems within the molecular dynamics framework.
The modification generalizes the previously developed linear correction to EAM-type potentials
[Sushko et al., J. Phys.: Condens. Matter \textbf{28}, 145201 (2016)]
and asymptotically approaches zero at large interatomic distances.
A general procedure for constructing this modification is outlined and its relation to the linear correction is elaborated.
To benchmark this procedure, we examine the melting phase transition and several equilibrium properties
of nanosystems made of silver, gold and titanium.
The simulations performed with the modified potential predict higher bulk melting temperatures of the metals
and agree better with experimental values as compared to the original EAM-type potential.
Our results show that the modification works well for metals with both cubic and hexagonal crystalline lattices.
The Gupta potential is chosen as an illustrative case study but the modification proposed is general and can be
applied to other widely-used potentials of the EAM type.
\end{abstract}

%\pacs{64.70.D-, 64.70.kd, 65.80.-g, 02.70.-c}

%\pacs{64.70.D- Solid-liquid transitions, 64.70.Nd Structural transitions in nanoscale materials, 64.70.kd Metals and alloys, 65.80.-g Thermal properties of small particles, nanocrystals, nanotubes, and other related systems, 02.70.-c Computational techniques; simulations}

% 64.70.D-   Solid-liquid transitions
% 64.70.kd   Metals and alloys
% 65.80.-g   Thermal properties of small particles, nanocrystals, nanotubes, and other related systems
% 02.70.-c   Computational techniques; simulations

%%%%%%%%%%%%%%%%%%%%%%%%%%%%%%%%%%%%%%%%%%%%%%%%%%%%%%%%%%%%%%%%%%%%%
%%%%%%%%%%%%%%%%%%%%%%%%%%%%%%%%%%%%%%%%%%%%%%%%%%%%%%%%%%%%%%%%%%%%%
\maketitle

\section{Introduction}

Computer simulations based on atomistic models have emerged as a powerful tool for the analysis
of physicochemical processes occurring in materials and related materials properties \cite{Handbook_MaterModel}.
A vast number of atomistic simulations employ the molecular dynamics (MD) method that requires evaluation of
the total potential energy of a many-atom system and the forces acting on constituent atoms
\cite{Rapaport_Art_of_MD, MBN_Explorer_Springer}.
MD simulations provide insights into many physical processes, such as
diffusion \cite{Hoyt_2000_PRL.85.594, Sushko_2014_JPhysChemA_diff, Cheng_2018_PRL.120.225901},
plastic deformation \cite{Verkhovtsev_2013_ComputMaterSci.76.20, Zink_2006_PRB.73.172203},
melting \cite{Cleveland_1998_PRL.81.2036, Qi_2001_JCP.115.385, Fang_2005_Nanotechnology.16.250, Zhang_2006_PRB.73.125443, Lyalin_2009_PRB.79.165403},
crystallization \cite{Qi_2001_JCP.115.385, Yakubovich_2013_PRB.88.035438}
and other phase transformations \cite{Mishin_2010_JPCM.22.295403, Kexel_2015_EPJB.88.221},
which all happen on the time and spatial scales exceeding by far those accessible by \textit{ab initio} methods.
To access these scales, semi-empirical interatomic potentials are used
\cite{MBN_Explorer_Springer,Kim_2009_JEngMaterTechnol.131.041210, Handbook_MD_potfunctions, Mueser_2015_MSMSE.23.070401},
which are parameterized for specific material compositions and structures. %, and certain temperature ranges.
An important issue is transferability of potentials \cite{Zhang_2018_SciRep.8.2424} -- a potential constructed by fitting to a
specific set of properties should perform well for other properties that were not considered during its construction phase.

Different interatomic potentials
\cite{TB-SMA_2, Sutton-Chen, Ackland_1992_PhilMagA.66.917, Cleri_1993_PRB.48.22, Foiles_1986_PRB.33.7983, Daw_1993_MaterSciRep.9.251} belonging to a general class of embedded-atom method (EAM) potentials are commonly used
in MD simulations of metal systems \cite{Mishin_2010_ActaMater.58.1117}.
Parameters of these potentials are derived to reproduce experimental data on the properties of bulk materials
(e.g., cohesive energy, equilibrium lattice constants, bulk modulus, elastic constants, vacancy-formation energy, etc.) or
they are fitted to reproduce those from zero-temperature \textit{ab initio} calculations of perfect crystalline structures.

It is also common that EAM-type potentials are less accurate in describing the dynamics
of systems being far from the equilibrium, for instance, the melting phase transition.
In particular, these potentials often cannot reproduce the experimental values of melting temperature
for bulk metals.
An illustrative example is titanium whose melting temperature calculated by means of different many-body potentials
deviates from the experimental value by several hundred degrees \cite{Kim_2006_PhysRevB.74.014101, Sushko_2014_JPhysChemA_diff}.
A similar level of discrepancy was observed for other metals,
e.g. gold \cite{Lewis_1997_PRB.56.2248, Ryu_2009_ModelSimulMaterSciEng.17.075008},
as well as for non-metal systems such as silicon \cite{Ryu_2009_ModelSimulMaterSciEng.17.075008}.
This indicates the necessity to modify the widely exploited force fields to achieve a more accurate description
of the systems' properties at elevated temperatures.
An accurate description of both equilibrium and non-equilibrium properties of metal systems is important,
for instance, for studying irradiation-driven phase and structural transformations of metal nanostructures \cite{Nordlund_1999_NIMB.159.183, Wang_2012_PRL.108.245502}
as well as irradiation-induced chemistry underlying novel nanofabrication techniques
\cite{Sushko_2016_EPJD.70.217, Huth_2012_BeilsteinJ.3.597}.

Different approaches to account for finite-temperature effects in classical force fields for metal systems
have been discussed in literature.
A method for re-parametrization of interaction potentials was proposed in Ref.~\citenum{Sturgeon_2000_PRB.62.14720}
to adjust the calculated melting temperature of materials without affecting the mechanical properties
to which the potentials were fitted.
%In that method, physical quantities describing the phase transition (the melting temperature or pressure)
%were calculated using a trial semi-empirical potential and
In that method, the melting temperature was calculated using a trial interatomic potential and
the Gibbs-Duhem equation (which relates changes in the chemical potential of a system to changes in its
temperature and pressure) was then solved to update parameters of the potential.
This method was applied \cite{Sturgeon_2000_PRB.62.14720} to re-parameterize an EAM-type potential for Al and
it improved the calculated bulk melting temperature without considerable change in other properties.
In a more recent work~\cite{Ackland_2012_JPCS.402.012001}, a correction to a many-body force field
for titanium was proposed which included the contribution of thermal excitations of electronic degrees of freedom.
In that approach, an EAM-type potential was augmented by an additional term (related to electronic entropy)
that arises from the Sommerfeld theory of metals.
According to the latter, there is a temperature-dependent contribution to the free energy of
a metal system which depends also on the density of states at the Fermi energy.
In Ref.~\citenum{Mendelev_2016_JCP.145.154102} several parameterizations of EAM-type potentials for Ti
describing defects, plasticity and melting were presented.
These potentials fit well to either low- or high-temperature experimental data
but could not describe both temperature regions simultaneously.
On this basis, a temperature-dependent potential, being a combination of the potentials operating better
in the different regions, was suggested to study the properties of Ti in a wide temperature range.
The knowledge accumulated in these studies suggests that the modification %or re-parametrization
of the conventional EAM-type potentials is required to bring the calculated non-equilibrium properties
(particularly, the melting temperature) of metal materials to the desired experimental values.

%A way to improve the description of non-equilibrium properties of materials is to adjust the functional form
%of a potential fitted to the ground-state properties without disturbing the latter.
%
In our previous work \cite{Sushko_2016_JPCM.28.145201} a modification of the widely-used Gupta potential \cite{Gupta}
was presented, which reproduced both the melting temperature and the near-equilibrium properties
of selected metal systems.
It was revealed that augmenting steepness of the interaction potential by enhancing
its repulsive part leads to an increase of the melting temperature.
This happens because the higher thermal energy is needed to reach the threshold of
atomic vibration amplitudes at which the melting occurs.
To that end, the original EAM-type Gupta potential was augmented by adding a
linear term to the repulsive part \cite{Sushko_2016_JPCM.28.145201}.
The linear correction represented a minor change to the potential energy but led to a significant increase of
the melting temperature.
It was applied to study the thermal, geometrical and energetic properties of magnesium, titanium, platinum and gold,
and a good agreement with experimental results was obtained.
In Ref. \citenum{Kexel_2016_JPCC.120.25043} this method was used to evaluate the melting points of
finite-size NiTi nanoalloys with different composition of Ni and Ti.
These results were used to evaluate bulk melting temperatures of Ni$_{1-x}$Ti$_{x}$ alloys,
which agreed with an experimental phase diagram for the NiTi material.

A drawback of the linear correction \cite{Sushko_2016_JPCM.28.145201} is its unphysical behavior
at large interatomic distances.
To avoid the continuous growth of the potential energy, interatomic interactions should be truncated
beyond a given cutoff distance, and the cutoff becomes another important parameter of the correction.

In this paper,
%we generalize the previously developed methodology and propose a
%novel modification of an EAM-type potential.
the previously developed methodology is generalized in the form of a new modification of an EAM-type potential.
% with the correct asymptotic behavior at large interatomic distances.
This modification represents a linear function multiplied by a sigmoid function which gradually tends to zero beyond a given distance.
A general procedure for constructing this modification is outlined and its parameters are
related to parameters of the linear correction \cite{Sushko_2016_JPCM.28.145201}.
The modified EAM-type potential is used for MD simulations of melting of nanometer-size
nanoparticles made of silver, gold and titanium.
Structural and energetic equilibrium properties of these systems, such as lattice constants, cohesive energy
and vacancy formation energy are also analyzed.
%The melting temperatures of $1-5$~nm gold nanoparticles calculated by means of the modified potential are
%in better agreement with experimental data \cite{Buffat_1975_PRA.13.2287} as compared to the original force field.
Our results demonstrate that the new modification is applicable for metals with both cubic and hexagonal crystalline lattices.
Similar to our previous works \cite{Sushko_2016_JPCM.28.145201, Kexel_2016_JPCC.120.25043}
the Gupta potential is chosen as an illustrative case study but the modification proposed can also be applied
to other widely-used interatomic potentials of the EAM type,
such as Sutton-Chen \cite{Sutton-Chen} or Finnis-Sinclair \cite{Finnis-Sinclair} potentials.

The paper organized as follows.
Section~\ref{sec:methods} describes the theoretical and computational approach.
In particular, the new modification to a EAM-type Gupta potential is introduced and
an analytical model is presented to derive its parameters for different metals.
In Section~\ref{sec:results} the modified potential is used to study the equilibrium properties
and melting of silver, gold and titanium crystals.
These results are compared with those obtained using the original Gupta potential
and the linear correction.
Finally, Section~\ref{sec:conclusions} summarizes the results of this work and gives
an outlook for further investigations.

%%%%%%%%%%%%%%%%%%%%%%%%%%%%%%%%%%%%%%%%%%%%%%%%%%%%%%%%%%%%%%%%%%%%%
%%%%%%%%%%%%%%%%%%%%%%%%%%%%%%%%%%%%%%%%%%%%%%%%%%%%%%%%%%%%%%%%%%%%%
\section{Theoretical and computational methodology}
\label{sec:methods}

\subsection{EAM-type Gupta potential}

As a case study, we consider the interatomic potential developed by Gupta \cite{Gupta}.
Similar to other many-body potentials of the EAM type
\cite{Daw_1993_MaterSciRep.9.251, Finnis-Sinclair, Sutton-Chen, Cleri_1993_PRB.48.22, TB-SMA_2},
%Daw_1983_PhysRevLett.50.1285, Daw_1984_PhysRevB.29.6443,
it is constructed as a sum of (i) a short-range repulsive term that stems
from the repulsion between atomic cores and (ii) a long-range attractive term
which imitates delocalization of the outer-shell electrons
and is related to electron density at a given atomic site.

The total energy of an $N$-atom system interacting via an EAM-type potential reads
\begin{equation}
U = \frac12 \sum_{i=1}^N  \sum_{j \ne i} V(r_{ij}) + \sum_{i=1}^N F_i(\rho_i)  \ .
\label{FS}
\end{equation}
Here $V(r_{ij})$ is the short-range repulsive interaction between atoms $i$ and $j$
separated by the distance $r_{ij}$.
The attractive term $F_i$ stands for the energy obtained by embedding an atom $i$ into
the local electron density $\rho_i$ provided by the remaining atoms of the system.
The functional form of $F_i(\rho_i)$ may vary in different EAM-type potentials
\cite{Mishin_2010_ActaMater.58.1117}
while the Gupta potential employs a specific form of this function, $F_i(\rho_i) \propto - \sqrt{\rho_i}$.
This functional form is based upon the second-moment approximation of the tight-binding model
\cite{Ackland_1988_JPF.18.L153, Goringe_1997_RepProgPhys.60.1447}, according to which the
attractive many-body term is related to the energy of the valence $d$-electron band and
expressed as a square root of $\rho_i$.
The latter is constructed empirically as a linear superposition of electron charge densities of
constituent atoms \cite{Daw_1983_PhysRevLett.50.1285, Finnis-Sinclair},
$\rho_i = \sum_{j \ne i} \varrho(r_{ij})$.

Within the Gupta representation the functions $V(r_{ij})$ and
$\varrho(r_{ij})$ are introduced in exponential forms so that the total potential energy $U_{\textrm{Gup}}$
reads as follows:
\begin{equation}
U_{\textrm{Gup}} =
\sum_{i=1}^N \left[ \frac12 \sum_{j \ne i}  A\,
{\rm e}^{-p \left( \frac{r_{ij}}{d} - 1 \right)} \right.
- \left. \sqrt{ \sum_{j \ne i} \xi^2\,
{\rm e}^{-2q \left( \frac{r_{ij}}{d} - 1 \right)}  } \, \right] \ .
\label{Finnis_Sinclair_pot}
\end{equation}
Here
$d$ is the first-neighbor distance,
%$p$ and $q$ control the decay of the exponential functions and 
$p$ and $q$ are related to bulk elastic constants \cite{Tomanek_1985_PhysRevB.32.5051},
$\xi$ represents an effective orbital-overlap integral,
and $A$ adjusts the cohesive energy.
The parameters for silver, gold and titanium used in this work are summarized in Table \ref{Table_FF_parameters}.

\begin{table}
\caption{Parameters of the Gupta potential describing interactions in silver, gold and titanium \cite{Cleri_1993_PRB.48.22}.}
\begin{tabular}{p{0.8cm}p{1.3cm}p{1.3cm}p{1.3cm}p{1.3cm}p{1.0cm}}
\hline
     & $d$~(\AA) & $A$~(eV) &    $p$   &  $\xi$~(eV) &   $q$   \\
\hline
 Ag &    2.889   &  0.1028  &  10.928  &    1.178    &  3.139  \\
 Au &    2.884   &  0.2061  &  10.229  &    1.790    &  4.036  \\
 Ti &    2.950   &  0.1519  &  \hspace{1mm}  8.620  &    1.811    &  2.390  \\
\hline
\end{tabular}
\label{Table_FF_parameters}
\end{table}

%%%%%%%%%%%
\subsection{Linear correction to EAM-type potentials}
\label{sec:linear-correction}

The Gupta potential (\ref{Finnis_Sinclair_pot}) corrected with the linear term $U_{\textrm{lin}}$
introduced in Ref.~\citenum{Sushko_2016_JPCM.28.145201} reads
\begin{subequations}
\begin{eqnarray}
U &=& U_{\textrm{Gup}} + U_{\textrm{lin}} \ ,
\label{Finnis_Sinclair_pot_mod_add} \\
U_{\textrm{lin}}  &=&  \frac12 \sum_{i,j=1}^N \left( B \, r_{ij} + C \right) \ ,
%&& U = U_{\textrm{Gup}} + U_{\textrm{lin}} ; \nonumber \\
%&& U_{\textrm{lin}} = \sum_{i=1}^N B \, r_{ij} + C \ ,
\label{Finnis_Sinclair_pot_mod}
\end{eqnarray}
\end{subequations}
where $B$ and $C$ are parameters.
The linear form ensures that the curvature of the modified potential energy profile
in the vicinity of the equilibrium point (governed by the second derivative of
potential energy $U$) coincides with that of the original Gupta potential.
This condition was set to leave intact near-equilibrium properties.

As discussed in Ref.~\citenum{Sushko_2016_JPCM.28.145201}, the term $B r_{ij}$ ($B > 0$) makes
the potential energy profile steeper at interatomic distances exceeding the equilibrium point $r_0$
whilst slightly changing the depth of the potential well at $r_0$.
The constant term $C < 0$ was added to mitigate the latter effect.
In Ref.~\citenum{Sushko_2016_JPCM.28.145201} the parameters $B$ and $C$ were obtained empirically
for a specific cutoff distance $r_c$ for titanium, gold, platinum and magnesium.
These parameters can be derived for any material and any $r_c$ using the following analytical estimate.

\begin{figure*}[htb!]
\centering
\includegraphics[width=0.9\textwidth,clip]{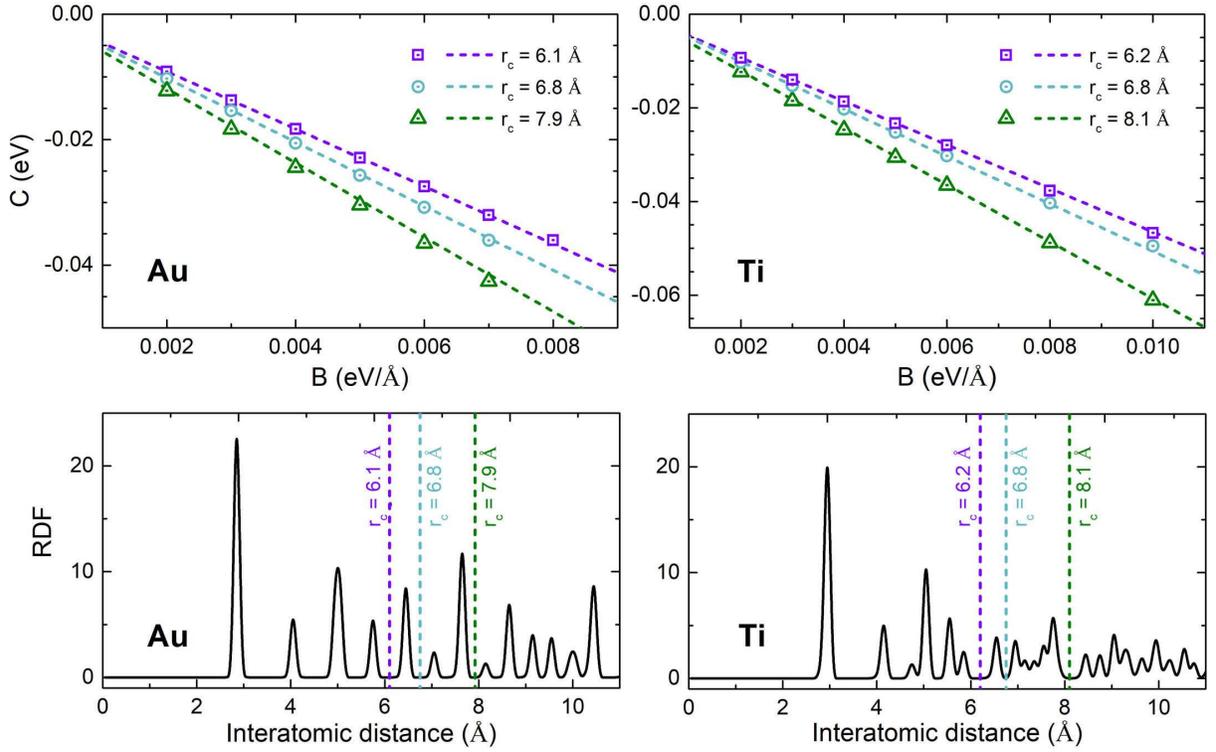}
\caption{Upper row: Functional relation between the parameters of the linear correction, $C(B)$,
which leave intact the cohesive energy of bulk metal systems.
Lines show the dependencies calculated using Eq.~(\ref{Lin_cond1}) for different cutoff values $r_c$
while symbols show the results of structure optimization calculations that account for the realistic
%(fcc for Au and hcp for Ti)
crystal structures (see the discussion in Section~\ref{sec:results_contmedium}).
Bottom row: Radial distribution functions (RDFs) for %the optimized
10-nm~nanoparticles (composed of approximately 30,000 atoms) made of gold and titanium.
The cutoff values used in the calculations are shown by dashed lines.}
\label{figure_B-C_linear}
\end{figure*}

The correction to an EAM-type potential should not change the cohesive energy of a bulk material
to which the potential was fitted.
Therefore, the change in total potential energy due to the linear correction should be equal to zero.
If we approximate the real crystalline structure of a metal with a uniform distribution of atoms
with number density $n_0$, this condition can be written as
\begin{equation}
%\Delta U = 
\int_{r < r_c} n_0 \, \left( B \, r + C \right) \, \textrm{d}V = 0 \ ,
\label{Lin_cond1_general}
\end{equation}
leading to the relation
%Then, the following condition is derived from Eq.~(\ref{Lin_cond1_general}):
\begin{equation}
%B \, \frac{r_c^4}{4} + C \, \frac{r_c^3}{3} = 0 \ \ \to \ \
%r_c = - \frac{4}{3} \, \frac{C}{B} \ .
C = - \frac{3}{4} \, B \, r_c  \ .
\label{Lin_cond1}
\end{equation}
The upper panels of Fig.~\ref{figure_B-C_linear} show by lines the calculated dependence $C(B)$
for gold and titanium for different values of $r_c$.
Bulk gold and silver have an fcc crystal lattice and very similar lattice constants,
so the results shown for gold describe also silver crystals.
For each metal we consider three cutoff distances between 6 and 8~\AA, corresponding to minima
in the radial distribution function (see the vertical lines in the bottom panels of Fig.~\ref{figure_B-C_linear}).
The indicated values of $r_c$ were chosen following Ref.~\citenum{Cleri_1993_PRB.48.22}.
In that work, the parameters of the Gupta potential for the fcc metals %(gold and silver)
were derived accounting for interatomic interactions up to the fifth-neighbor shell, while
the suggested cutoff values for titanium and other hcp structures
corresponded to inclusion of 7-8 shells of neighboring atoms.

%which approximates
%the real crystalline structure of metals with a uniform distribution of atoms.
%As shown below, the values of $B$ and $C$ obtained with this model
%give good agreement between the calculated and experimental values of the cohesive energy
%and the melting temperature of metals. Details of this model are given below.
%
%Therefore, the contribution to the total potential energy of a system due to $U_{\textrm {lin}}$ should be equal to zero.
%
%where $n(r)$ is the atomic density of a system. % and $r_c$ is the cutoff distance.
%set to avoid non-physical effects arising from the interactions at larger interatomic distances.
%Let us approximate it by a uniform distribution of atoms, $n(r) = n_0 = \rho \, N_\textrm{A}/M$,
%where $\rho$ is the mass density of a material, $M$ its molar mass, and $N_\textrm{A}$ the Avogadro's number.
%The parameters for the three metals considered are summarized in Table~\ref{tab:density_param}.
%

%The linear correction causes a small displacement of atoms from their equilibrium positions defined by
%the original Gupta potential.
%A change in potential energy associated with the displacement of an atom by $\Delta r$ is given by
%
The potential energy $U$ can be expanded in a Taylor series about the equilibrium atomic positions
for the original Gupta potential.
Keeping only the first term of the expansion one can evaluate the change in potential energy associated
with the displacement of an atom by $\Delta r$ due to the linear correction:
\begin{equation}
\Delta U
= - F_{\textrm{lin}} \, \Delta r
= - \frac{2\pi}{3} \, \left( \frac{4}{3} \right)^3 \, \frac{C^3}{B^2} \, n_0 \, \Delta r \ .
\label{eq:linear_condition2}
\end{equation}
%
%\begin{equation}
%\frac{\Delta U}{\Delta r} = \frac{2\pi}{3} \, \left( \frac{4}{3} \right)^3 \, \frac{C^3}{B^2} \, n_0
%\end{equation}
%
%where
%$F_{\textrm{lin}} = \int_{r < r_c} n_0 \left( - \frac{\textrm{d}U_{\textrm{lin}}}{\textrm{d}r} \right) \textrm{d}V$
%and $r_c$ is derived from Eq.~(\ref{Lin_cond1}).

%The change in potential energy is related
%to the change in thermal energy of the system, $\Delta U = k_{\textrm B}\Delta T$,
%where $k_{\textrm B}$ is the Boltzmann constant.
%Introduction of the linear correction increases the average amplitude of thermal vibrations.
%This requires a larger amount of thermal energy to initiate the phase transition and thus
%leads to an increase of the melting temperature.
%Knowing the experimental bulk melting temperature $T_{\textrm m}^{\textrm{exp}}$ and the value
%predicted by the Gupta potential, $T_{\textrm m}^{\textrm{Gup}}$, one can evaluate
%the force $F_{\textrm{lin}}$ and the displacement $\Delta r$ needed
%to increase the melting temperature by $\Delta T = T_{\textrm m}^{\textrm{exp}} - T_{\textrm m}^{\textrm{Gup}}$.
%Details of this analysis for gold, silver and titanium are presented in Section~\ref{sec:results_contmedium}.

As it was demonstrated in our earlier work~\cite{Sushko_2016_JPCM.28.145201}, augmenting steepness of the interatomic
potential beyond the equilibrium point by enhancing the repulsive contribution of the force field leads to a rise
of the melting point.
It happens because the increased thermal energy is needed to reach the threshold of atomic vibration amplitudes
at which the melting phase transition occurs.
Knowing the experimental bulk melting temperature $T_{\textrm m}^{\textrm{exp}}$ and the value predicted
by the Gupta potential, $T_{\textrm m}^{\textrm{Gup}}$, the parameters $B$ and $C$ can be chosen such that
the melting temperature will increase by $\Delta T = T_{\textrm m}^{\textrm{exp}} - T_{\textrm m}^{\textrm{Gup}}$.

Conditions (\ref{Lin_cond1}) and (\ref{eq:linear_condition2}) define,
for any $r_c$, a unique set of parameters $(B,C)$ that reproduce experimental values of
cohesive energy and melting temperature of bulk materials.
These conditions were used to define $B$ and $C$ for the three metals studied.

\subsection{Generalized modification of EAM-type potentials}
\label{sec:new-modification}

\begin{figure}[htb!]
\centering
\includegraphics[width=0.48\textwidth,clip]{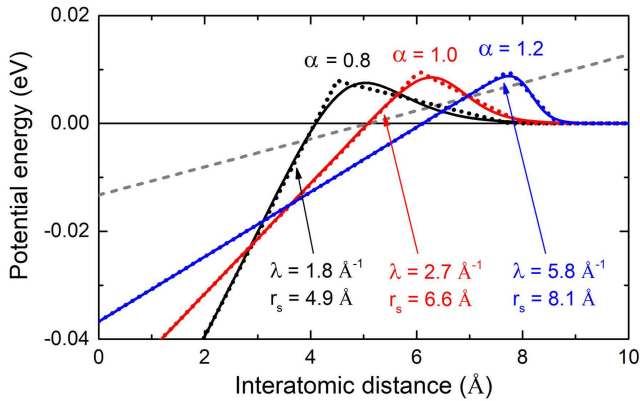}
\caption{
%Potential energy of a silver dimer interacting via $U_{\textrm{mod}}$ (\ref{eq:newpot}).
Potential energy $U_{\textrm{mod}}$, Eq.~(\ref{eq:newpot}), as a function of interatomic distance $r$.
Solid lines show $U_{\textrm{mod}}(r)$ for different values of the parameters,
which were derived using the procedure described in the text.
Dotted curves show the piecewise linear approximation $\bar{U}_{\textrm{mod}}(r)$, Eq.~(\ref{eq:newpot_approximation}),
obtained for different values of the parameter $\alpha$
(see Appendix \ref{sec_appendix} for details).
Dashed gray line depicts the potential $U_{\textrm{lin}}$.
% that predicts the correct melting temperature for silver
%with the cutoff $r_c = 6.8$~\AA~($B = 0.0026$~eV/\AA, $C = -0.013$~eV).
}
\label{figure_newpot_alphas}
\end{figure}

In this Section, we generalize the above described methodology and propose a new modification
of an EAM-type potential.
It should keep features of the linear correction, i.e. maintain its behavior in the vicinity
of atomic equilibrium points and enhance the repulsive interactions with the growth of atomic displacements.
%
%Importantly, it should also be free of the non-physical behavior at large interatomic distances.
The modification should also contain an additional parameter describing the characteristic range of
the potential thus eliminating the dependence of the potential on the choice of cutoff distance.
These conditions are fulfilled by multiplying the linear correction (\ref{Finnis_Sinclair_pot_mod})
by a sigmoid function, which is equal to unity at small interatomic distances and asymptotically
approaches zero beyond a given distance.
The modified Gupta potential then reads as
\begin{subequations}
\begin{eqnarray}
U &=& U_{\textrm{Gup}} + U_{\textrm{mod}} \ ,
\label{eq:newpot_add} \\
U_{\textrm{mod}} &=& \frac12 \sum_{i,j=1}^N \frac{\tilde{B} \, r_{ij} + \tilde{C}}{1 + e^{\lambda (r_{ij} - r_s)}} \ .
%&& U(r_{ij}) = U_{\textrm{Gup}}(r_{ij}) + U_{\textrm{mod}}(r_{ij}) ; \nonumber \\
%&& U_{\textrm{mod}}(r_{ij}) = \frac{\tilde{B} \, r_{ij} + \tilde{C}}{1 + e^{\lambda (r_{ij} - r_s)}} \ .
\label{eq:newpot}
\end{eqnarray}
\end{subequations}
The parameters $\tilde{B}$ and $\tilde{C}$ have the same meaning as $B$ and $C$ in Eq.~(\ref{Finnis_Sinclair_pot_mod}):
$\tilde{B}$ defines the additional force acting on the nearest atoms
and $\tilde{C}$ adjusts the depth of the potential well in the vicinity of the equilibrium point where $U = 0$.
The parameter $\lambda$ describes the slope of $U_{\textrm{mod}}$ at large interatomic distances,
while $r_s$ defines the sigmoid's midpoint and hence the range of this potential.
Figure~\ref{figure_newpot_alphas} shows the potential $U_{\textrm{mod}}$ for a pair of atoms
as a function of interatomic distance $r$.
Due to its sigmoid-type shape, $U_{\textrm{mod}}(r)$ asymptotically approaches zero and
its range serves as a natural cutoff distance for this interaction.
%Varying the parameters of the potential, its range can be extended towards large interatomic distances.

For each pair of atoms the potential $U_{\textrm{lin}}$ grows monotonically with interatomic distance
up to the cutoff $r_c$,
and different shells of neighboring atoms located within the sphere of radius $r_c$ experience
a constant force exerted by a given atom.
On the contrary, $U_{\textrm{mod}}$ has a maximum at interatomic distances of about 5--8~\AA~depending
on the choice of $\lambda$ and $r_s$ (see Fig.~\ref{figure_newpot_alphas}).
Thus, the force exerted by an atom due to $U_{\textrm{mod}}$ enhances interaction with several nearest
atomic shells while the interaction with more distant atoms weakens.
The strength of this interaction is governed by steepness of the potential beyond the maximum,
i.e., by the parameter $\lambda$.
Therefore, the force acting on the nearest neighbors due to $U_{\textrm{mod}}$ should exceed (by the absolute value)
the force $F_{\textrm{lin}}$ as its effect is compensated by the weaker interaction with more distant atoms.
This means that for each pair of atoms interacting via $U_{\textrm{mod}}(r)$
the initial slope of the potential
should be steeper than the slope of $U_{\textrm{lin}}(r)$, i.e., $\tilde{B} > B$.

To analytically derive parameters of the new modification, %can be
$U_{\textrm{lin}}(r)$ in Eqs.~(\ref{Lin_cond1_general}) and (\ref{eq:linear_condition2})
 %should be
%was substituted with $U_{\textrm{mod}}(r)$
%However, to the best of our knowledge, such integrals are not listed in the well-known mathematical
%handbooks \cite{Prudnikov_integrals_Vol1} and cannot be done analytically.
%To derive the parameters of $U_{\textrm{mod}}(r)$, it was approximated
%by a piecewise linear function $\bar{U}_{\textrm{mod}}(r)$, see Eq.~(\ref{eq:newpot_approximation}) in the Appendix.
%
was substituted with $\bar{U}_{\textrm{mod}}(r)$, -- a piecewise linear approximation of the sigmoid-type
function $U_{\textrm{mod}}(r)$, see Eq.~(\ref{eq:newpot_approximation}) in Appendix.
Then, parameters of this function were expressed through the parameters $B$ and $C$ of the linear correction.
As the last step of this procedure, $\bar{U}_{\textrm{mod}}(r)$ was fitted with $U_{\textrm{mod}}(r)$ to derive
$\lambda$ and $r_s$.
%the parameters $\lambda$ and $r_s$ of the latter function.
Further technical details are given in Appendix.
%
%Section~\ref{sec:results} presents results of the analysis of the melting temperature and
%several near-equilibrium properties of finite-size and bulk metal systems,
%evaluated by means of the modified Gupta potential (\ref{eq:newpot_add})--(\ref{eq:newpot}).
%
The parameters of $U_{\textrm{mod}}$ %and $\bar{U}_{\textrm{mod}}(r)$ which were
used for the analysis of the melting temperature and near-equilibrium properties of
silver, gold and titanium systems are summarized in Table~\ref{Table_Umod_parameters}.
Details of this analysis are presented below in Section~\ref{sec:results_Umod}.

\begin{table}
\caption{Parameters of the potential $U_{\textrm{mod}}$, Eq.~(\ref{eq:newpot}),
used to analyze the melting temperature and equilibrium properties of silver, gold and titanium.}
%\begin{tabular}{p{0.8cm}p{2.0cm}p{2.0cm}p{1.7cm}p{1.7cm}p{1.0cm}}
%\hline
%    & $\tilde{B}$~(eV/\AA) & $\tilde{C}$~(eV) & $\lambda$~(\AA$^{-1}$) & $r_s$~(\AA) & $R_2$~(\AA) \\
%\hline
% Au &          0.026       &       -0.145     &         4.681          &    7.358    &    8.056    \\
% Ag &          0.009       &       -0.048     &         5.933          &    7.098    &    7.628    \\
% Ti &          0.052       &       -0.269     &         2.765          &    6.681    &    8.065    \\
%\hline
%\end{tabular}
%
\begin{tabular}{p{0.9cm}p{1.7cm}p{1.7cm}p{1.7cm}p{1.3cm}}
\hline
    & $\tilde{B}$~(eV/\AA) & $\tilde{C}$~(eV) & $\lambda$~(\AA$^{-1}$) & $r_s$~(\AA)  \\
\hline
 Ag &          0.009       &       -0.048     &         5.933          &    7.098     \\
 Au &          0.026       &       -0.145     &         4.681          &    7.358     \\
 Ti &          0.052       &       -0.269     &         2.765          &    6.681     \\
\hline
\end{tabular}
\label{Table_Umod_parameters}
\end{table}

The modification $U_{\textrm{mod}}$ (\ref{eq:newpot}) is qualitatively similar to the well-known Dzugutov potential \cite{Dzugutov_1992_PRA.46.2984} which was developed to model glass-forming liquid metals.
The Dzugutov potential coincides with the Lennard-Jones potential at small interatomic distances
but has a maximum beyond the equilibrium point.
The position of the maximum is between the first- and the second-neighbor shells in icosahedral structures and,
at the same time, it corresponds to distances characteristic for short-range crystalline order.
This enables the suppression of crystallization and enforces the emergence of icosahedral structures.
The maximum of $U_{\textrm{mod}}$ corresponds to the positions of more distant atoms located
in the fifth to ninth neighboring shells (see Fig.~\ref{figure_newpot_alphas} and the RDFs in Fig.~\ref{figure_B-C_linear}).
As a result, it does not change crystal structure but enables an increase of the melting temperature
whilst slightly changing the near-equilibrium properties of metals.

\subsection{Computational details}

All simulations described in this work were conducted using MBN Explorer \cite{MBN_Explorer1}
-- a software package for advanced multiscale modeling of complex molecular structure and dynamics,
equipped with a large library of pairwise and many-body potentials \cite{MBN_Explorer_Springer}.
Spherical nanoparticles with radii from 1 to 5~nm (ranging from 250 to 30,000 atoms), cut from ideal
silver, gold and titanium crystals, were constructed by means of the MBN Studio software \cite{MBN_Studio, MBN_Studio_Tutorials}.

Prior the analysis of the structural and energetic parameters of the systems
(lattice constants, cohesive energy and energy of vacancy formation)
energy minimization calculations were performed using the velocity-quenching algorithm.
The MD simulations of the melting process were performed using a large simulation box of
$20 \times 20 \times 20$~nm$^3$ in the NVT canonical ensemble.
The temperature $T$ was controlled by a Langevin thermostat with a damping time of 1~ps.
The nanoparticles were heated up (starting from the initial temperature $T_0$ well below
the expected melting temperatures, $T_0 = 300$~K for Ag and Au and 1000~K for Ti)
with a constant heat rate of 0.5~K/ps, which is within the range of typical values used for MD simulations
of phase transitions.
The total simulation time for each run was 3~ns.
The time integration of the equations of motion was done using the velocity-Verlet algorithm \cite{Rapaport_Art_of_MD}
with an integration time step of 1~fs.
In the calculations performed with the linear correction $U_{\textrm{lin}}$, the interatomic interactions
were truncated at the cutoff radius $r_c$ ranging from about 6 to 8~\AA.
In the case of the potential augmented with $U_{\textrm{mod}}$ the range of the latter
served as a natural cutoff distance, which varied between 8 and 9~\AA.

Melting temperature of the nanoparticles was determined from the analysis of heat capacity,
$C_V = \left( \partial E / \partial T \right)_V$, defined as a partial derivative of
the internal energy of the system with respect to temperature at a given volume.
A sharp maximum of $C_V$ was attributed to the nanoparticle melting and the position of the maximum was referred to as
the nanoparticle melting point.
The bulk melting temperature was estimated by extrapolating the obtained
values to the bulk ($N \to \infty$) limit according to the Pawlow law\cite{Pawlow_1909_ZPhysChem.65, Calvo_2015_PCCP.17.27922}.
It describes the dependence of the melting temperature of spherical particles
on the number of atoms they are composed of as
$T_{\textrm{m}} = T_{\textrm{m}}^{\rm bulk} - \gamma N^{-1/3}$, where $T_{\textrm{m}}^{\rm bulk}$ is the
melting temperature of a bulk material and $\gamma$ is the factor of proportionality.

\section{Results and discussion}
\label{sec:results}

\subsection{Validity of the uniform density model}
\label{sec:results_contmedium}

The upper panel of Fig.~\ref{figure_B-C_linear} shows the dependence $C(B)$ that describes
the parameters of the linear correction $U_{\textrm{lin}}$ at different values of cutoff $r_c$.
Dashed lines were obtained by means of Eq.~(\ref{Lin_cond1}) within the uniform density model
(see Sect.~\ref{sec:linear-correction}),
while symbols show the results of structure optimization of gold and titanium systems with the
realistic (fcc and hcp, respectively) crystal structures.
In the case of structure optimization the parameters $B$ and $C$ were chosen to match experimental cohesive energies \cite{Kittel}.
The outcomes of the uniform density model agree nicely with the results of optimization calculations.
Table \ref{tab:cohesive-energy} summarizes the bulk cohesive energy for silver, gold and titanium,
calculated with the linear correction as well as the experimental values and the results obtained
by means of the original Gupta potential.

\begin{table}
\centering
\caption{Bulk cohesive energy (in eV per atom) calculated with the original Gupta potential, Eq.~(\ref{Finnis_Sinclair_pot}),
as well as with the Gupta potential corrected by $U_{\textrm{lin}}$, Eq.~(\ref{Finnis_Sinclair_pot_mod}),
and by the sigmoid-type modification $U_{\textrm{mod}}$, Eq.~(\ref{eq:newpot}), proposed in this work.
Experimental values are taken from Ref.~\citenum{Kittel}. }
\begin{tabular}{p{0.9cm}p{1.5cm}p{2.1cm}p{2.3cm}p{1.0cm}}
%\begin{tabular}{p{1.0cm}p{1.8cm}p{2.5cm}p{2.7cm}p{1.1cm}}
\hline
    &  $U_{\textrm{Gup}}$  &  $U_{\textrm{Gup}}$ + $U_{\textrm{lin}}$  &  $U_{\textrm{Gup}}$ + $U_{\textrm{mod}}$  &  exp.   \\
\hline
Ag  &        2.96       &         2.96      &   2.97    &   2.96   \\
Au  &        3.78       &         3.77      &   3.78    &   3.78   \\
Ti  &        4.87       &         4.87      &   4.83    &   4.85   \\
\hline
\end{tabular}
\label{tab:cohesive-energy}
\end{table}

\begin{figure*}[htb!]
%\centering
%\hspace{-1.0cm}
\includegraphics[width=1.0\textwidth,clip]{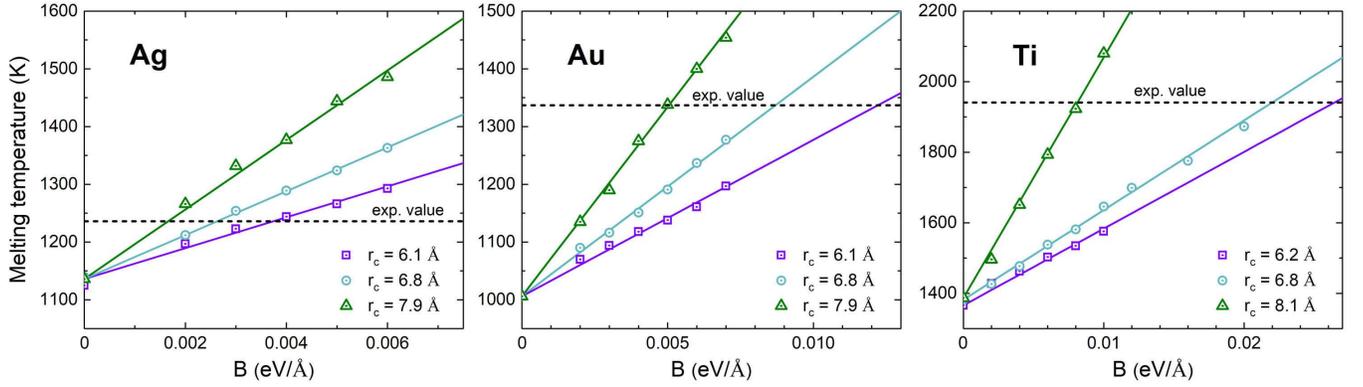}
\caption{Melting temperature of bulk silver, gold and titanium calculated using the linear correction $U_{\textrm{lin}}$
to the Gupta potential, Eq.~(\ref{Finnis_Sinclair_pot_mod_add})--(\ref{Finnis_Sinclair_pot_mod}),
at different values of the parameter $B$ and the cutoff $r_c$.
Symbols denote the results of MD simulations of finite-size nanoparticles melting, extrapolated to the bulk limit.
Lines represent the least-squares fit to these results.
$B = 0$ corresponds to the original Gupta potential, Eq.~(\ref{Finnis_Sinclair_pot}).
Experimental values from Ref.~\citenum{Kittel} are shown by dashed lines.}
\label{figure_Tmelt_linear}
\end{figure*}

Figure~\ref{figure_Tmelt_linear} shows the bulk melting temperature for silver, gold and titanium
calculated using the Gupta potential corrected by $U_{\textrm{lin}}$.
Different values of $B$ and $r_c$ were considered and the parameter $C$ was defined according to Eq.~(\ref{Lin_cond1}).
The figure shows that the calculated melting temperature increases linearly with $B$.
These results can be used to evaluate $\Delta T(B) = T_{\textrm m}^{\textrm{lin}}(B) - T_{\textrm m}^{\textrm{Gup}}$.
As follows from Eq.~(\ref{eq:linear_condition2}), $\Delta U = k_{\textrm{B}} \Delta T  \propto \Delta r$,
where $\Delta r$ stands for an increase in the amplitude of thermal vibrations of atoms with respect to the values
predicted by the original Gupta potential.
The slope of $\Delta T(B)$ is therefore proportional to the distance by which the atoms should be
additionally displaced from equilibrium positions to initiate the melting process at the temperature
corresponding to the experimental value.
For silver and gold $\Delta r \approx 0.09$~\AA, which is about 3\% of the nearest-neighbor distances,
$d_{\textrm{Ag}} = 2.889$~\AA~and $d_{\textrm{Au}} = 2.884$~\AA.
For titanium we observed the dependence of $\Delta r$ on the cutoff distance. 
For smaller cutoff values, $r_c = 6.2$~\AA~and 6.8~\AA,
the increase in the amplitude of thermal vibrations is equal to 0.06~\AA~and it increases up to 0.09~\AA~for $r_c = 8.1$~\AA.
These results suggest that an increase in the amplitude of thermal vibrations by a few percent leads
to a dramatic increase in the melting point.
A much steeper slope of $\Delta T(B)$ for titanium at $r_c = 8.1$~\AA~suggests that the distant atoms
located in a concentric shell between 7~\AA~and 8~\AA~make a significant contribution to the melting process and the
uncorrected Gupta potential cannot account properly for this contribution.

\subsection{Benchmarking the modified potential}
\label{sec:results_Umod}

Tables \ref{tab:cohesive-energy}--\ref{tab:lattice-const} summarize the results on structural and
energetic properties of silver, gold and titanium crystals obtained with the sigmoid-type modification
of the Gupta potential ({\ref{eq:newpot_add})--(\ref{eq:newpot}).
%A generality of the introduced modification is emphasized by considering metals
%with different (fcc and hcp) crystal lattice structures.
The results obtained with $U_{\textrm{mod}}$ are compared to those obtained
by means of the original Gupta potential (\ref{Finnis_Sinclair_pot}) and the linear correction
$U_{\textrm{lin}}$ (\ref{Finnis_Sinclair_pot_mod_add})--(\ref{Finnis_Sinclair_pot_mod}).

\begin{table*}
\centering
\caption{Vacancy formation energy $E_{\rm vf}$ (in eV) calculated with the original Gupta potential ($U_{\textrm{Gup}}$),
as well as with the Gupta potential corrected by $U_{\textrm{lin}}$ and the new modification $U_{\textrm{mod}}$.
Experimental values (``exp.'') as well as the results of MD simulations performed with different EAM-type potentials
and DFT calculations (``calc.'') are also indicated for comparison. FS stands for the Finnis-Sinclair potential and MEAM is modified EAM. }
%\begin{tabular}{p{0.9cm}p{1.4cm}p{1.7cm}p{2.0cm}p{2.1cm}p{2.5cm}}
\begin{tabular}{p{0.9cm}p{1.4cm}p{2.5cm}p{2.7cm}p{2.7cm}p{2.5cm}p{1.0cm}}
\hline
    &  $U_{\textrm{Gup}}$  &  $U_{\textrm{Gup}}$ + $U_{\textrm{lin}}$  &  $U_{\textrm{Gup}}$ + $U_{\textrm{mod}}$   &  exp.  &  \qquad \qquad \quad calc. & \\ \cline{6-7}
    &                     &                                           &                                           &        & EAM-type & DFT \\
\hline
%Ag  &  0.94   &    0.90   &   0.91   & $0.99 \pm 0.06$ \cite{McGervey_1973_PLA.44.53}        &  0.78 \cite{TB-SMA_2} (TB) \\
%    &         &           &                     & $1.09 \pm 0.10$ \cite{Simmons_1960_PhysRev.119.600}   &  0.88 \cite{Cleri_1993_PRB.48.22} (TB) \\
Ag  &  0.94   &    0.90   &   0.91   & $0.99 \pm 0.06$ \cite{McGervey_1973_PLA.44.53}        &  0.78 \cite{TB-SMA_2} (Gupta) & \\
    &         &           &                     & $1.09 \pm 0.10$ \cite{Simmons_1960_PhysRev.119.600}   &  0.88 \cite{Cleri_1993_PRB.48.22} (Gupta) & \\
    &         &           &                     &                                &  0.97 \cite{Foiles_1986_PRB.33.7983} (EAM) & \\
    &         &           &                     &                                &  1.10 \cite{Doyama_1997_RadEffDefSol.142.107} & \\
\hline

%Au  &  0.72   &    0.58   &  0.81  &  $0.62 - 0.67$ \cite{Jongenburger_1957_PhysRev.106.66}  &  0.60 \cite{TB-SMA_2} (TB) \\
%    &         &           &                   &  $0.70 - 1.10$ \cite{Jongenburger_1957_PhysRev.106.66}  &  0.75 \cite{Cleri_1993_PRB.48.22} (TB) \\
Au  &  0.72   &    0.58   &  0.81  &  $0.62 - 0.67$ \cite{Jongenburger_1957_PhysRev.106.66}  &  0.60 \cite{TB-SMA_2} (Gupta) & \\
    &         &           &                   &  $0.70 - 1.10$ \cite{Jongenburger_1957_PhysRev.106.66}  &  0.75 \cite{Cleri_1993_PRB.48.22} (Gupta) & \\
    &         &           &                   &                                  &  1.03 \cite{Foiles_1986_PRB.33.7983} (EAM) & \\
    &         &           &                   &                                  &  1.01 \cite{Doyama_1997_RadEffDefSol.142.107} & \\
\hline
%Ti  &  1.49   &   1.22    &    1.44                &   1.55 \cite{Shestopal_1966_FTT.7.3461}  & 1.56 \cite{Lai_2000_JPCM.12.L53} (TB) \\
%    &         &           &                        &                             & 1.43 \cite{Ackland_1992_PhilMagA.66.917} (TB) \\
Ti  &  1.49   &   1.22    &    1.44                &   1.55 \cite{Shestopal_1966_FTT.7.3461}  & 1.43 \cite{Ackland_1992_PhilMagA.66.917} (FS) & 1.97 \cite{Raji_2009_PhilosMag.89.1629} \\
    &         &           &                        &                             & 1.49 \cite{Johnson_1991_PhilMagA.63.865} (EAM)  & 2.14 \cite{LeBacq_1999_PRB.59.8508} \\
    &         &           &                        &                             & 1.56 \cite{Lai_2000_JPCM.12.L53} (Gupta) & \\
    &         &           &                        &                             & 1.78 \cite{Baskes_1994_MSMSE.2.147} (MEAM) & \\
    &         &           &                        &                             & 1.79 \cite{Kim_2006_PhysRevB.74.014101} (MEAM) & \\
\hline
\end{tabular}
\label{tab:vacancy-energy}
\end{table*}

% Ag - Rosato 1989: Gupta
%      Cleri-Rosato 1993: Gupta
%      Foiles 1986: EAM

Table~\ref{tab:cohesive-energy} presents the bulk cohesive energy.
Both the linear correction and the new sigmoid-type modification almost do not change the values predicted by the
original Gupta potential, and all these values are in good agreement with experimental data \cite{Kittel} with the relative
discrepancy of less than 0.5\%.

Table~\ref{tab:vacancy-energy} presents the vacancy-formation energy defined as the amount of cohesive energy
required to form a vacancy in a perfect crystal.
The potential energies of systems containing $N$ and $N-1$ atoms read as $E_N = N E^{\rm coh}_N$
and $E_{N-1} = (N-1) \, E^{\rm coh}_{N-1}$, respectively,
where $E^{\rm coh}_N$ and $E^{\rm coh}_{N-1}$ are the corresponding cohesive energies per atom.
The vacancy-formation energy is then defined as \cite{Korzhavyi_1999_PRB.59.11693, Mattsson_2002_PRB.66.214110}
$E_{\rm vf} \equiv (N-1) \, (E^{\rm coh}_{N-1} - E^{\rm coh}_{N}) =  E_{N-1} - \frac{N - 1}{N} \, E_N$.
%where $E_N$ is the energy of a perfect crystal structure containing $N$ atoms and
%$E_{N-1}$ is the energy of the relaxed structure with a vacancy.
The calculated values (columns labeled as ``$U_{\textrm{Gup}}$'', ``$U_{\textrm{Gup}}$ + $U_{\textrm{lin}}$''
and ``$U_{\textrm{Gup}}$ + $U_{\textrm{mod}}$'') are compared with available experimental data (``exp.'')
and the results of DFT calculations and MD simulations employing different EAM-type potentials (``calc.'').

The values calculated with the original Gupta potential are consistent with some experimental
and theoretical values reported in literature \cite{McGervey_1973_PLA.44.53, Foiles_1986_PRB.33.7983, Jongenburger_1957_PhysRev.106.66, Cleri_1993_PRB.48.22, Johnson_1991_PhilMagA.63.865}, whereas other works predicted either smaller or much larger values of $E_{\rm vf}$.
Note that the theoretical results reported in literature were obtained with different EAM-type potentials
(Finnis-Sinclair and Gupta potentials as well as a distinct potential introduced in Ref.~\citenum{Doyama_1997_RadEffDefSol.142.107})
as well as with tabulated EAM and modified EAM (MEAM) potentials.
The variety of potentials and parameterizations used has resulted in a large (up to 40\%) discrepancy
between the calculated values of $E_{\rm vf}$.

Calculations performed with the Gupta potential corrected by $U_{\textrm{lin}}$
(see the column ``$U_{\textrm{Gup}}$ + $U_{\textrm{lin}}$'') yield smaller values of $E_{\rm vf}$
as compared to the original Gupta potential, and the magnitude of the decrease depends on the parameter $B$.
The values of $E_{\rm vf}$ listed in Table~\ref{tab:vacancy-energy} were obtained for each metal using
the $B$ values that reproduce the experimental bulk melting temperatures, see Fig.~\ref{figure_Tmelt_linear}.
The figure shows that for $r_c \approx 8$~\AA~the value of $B$ for silver, 0.0016~eV/\AA,
is three times smaller than that for gold, 0.005~eV/\AA, and five times smaller than
for titanium, 0.008~eV/\AA.
As a result, the vacancy-formation energy for silver calculated by means of the linear correction is slightly (by about 5\%)
smaller than the value predicted by the original Gupta potential.
For gold and, especially, titanium, larger values of $B$ should be used to reproduce the experimental bulk melting temperatures,
which leads to a more pronounced (by about 20\%) decrease of $E_{\rm vf}$.
However, as presented in Table~\ref{tab:vacancy-energy}, the magnitude of this discrepancy for titanium is within the
uncertainty range of the existing theoretical data obtained by means of different EAM-type potentials.
In the MD simulations reported in literature \cite{Ackland_1992_PhilMagA.66.917, Lai_2000_JPCM.12.L53, Johnson_1991_PhilMagA.63.865, Baskes_1994_MSMSE.2.147, Kim_2006_PhysRevB.74.014101} $E_{\rm vf}$ varies from about 1.4 to 1.8 eV
%Calculations performed with different EAM-type potentials predicted the vacancy-formation energy for Ti
%in the range from about 1.4 to 1.8 eV,
while DFT calculations \cite{Raji_2009_PhilosMag.89.1629,LeBacq_1999_PRB.59.8508} predicted even larger values up to 2.1 eV.

The sigmoid-type modification $U_{\textrm{mod}}$ gives the results which are closer to the experimental values and
the results of other MD simulations \cite{Cleri_1993_PRB.48.22, Foiles_1986_PRB.33.7983, Ackland_1992_PhilMagA.66.917, Johnson_1991_PhilMagA.63.865}.
%As discussed above in Section \ref{sec:new-modification},
%the repulsive force arising at large interatomic distances due to $U_{\textrm{mod}}$ counterbalances the attractive force acting on the nearest atoms, which
This is due to the correction of the asymptotic behavior of the original Gupta potential, 
i.e. the weakening of interatomic interactions at large distances.
%leads to a better agreement with the data available in literature.

\begin{table}
\centering
\caption{Equilibrium lattice constants (in~\AA) calculated with the original Gupta potential ($U_{\textrm{Gup}}$),
as well as with the Gupta potential corrected by $U_{\textrm{lin}}$ and the new modification $U_{\textrm{mod}}$.
Two lattice parameters, $a$ and $c$, are listed for titanium.
Experimental values are taken from Ref.~\citenum{Kittel}. }
\begin{tabular}{p{0.9cm}p{1.5cm}p{2.1cm}p{2.3cm}p{1.0cm}}
%\begin{tabular}{p{1.5cm}p{1.9cm}p{2.5cm}p{2.7cm}p{1.1cm}}
\hline
    &  $U_{\textrm{Gup}}$  &  $U_{\textrm{Gup}}$ + $U_{\textrm{lin}}$  &  $U_{\textrm{Gup}}$ + $U_{\textrm{mod}}$  &  exp.   \\
\hline
Ag  &        4.07   &      4.05    &     4.07    &   4.09   \\
Au  &        4.06   &      4.03    &     4.09    &   4.08   \\      % ++
Ti ($a$) &   2.91   &      2.83    &     2.89    &   2.95   \\      % ++
Ti ($c$) &   4.75   &      4.63    &     4.77    &   4.68   \\      % ++
\hline
\end{tabular}
\label{tab:lattice-const}
\end{table}

\begin{figure*}[htb!]
%\centering
%\hspace{-1.0cm}
\includegraphics[width=1.0\textwidth,clip]{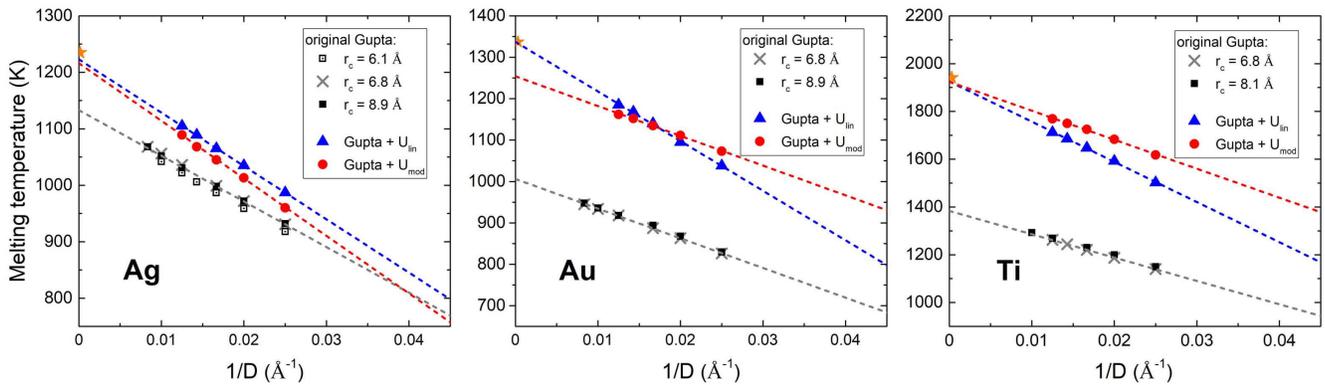}
\caption{
Melting temperature of Ag, Au and Ti nanoparticles of diameter $D$ calculated by means of the original Gupta potential
(Eq.~(\ref{Finnis_Sinclair_pot})), its linear correction $U_{\textrm{lin}}$ (Eq.~(\ref{Finnis_Sinclair_pot_mod}))
and the new modification $U_{\textrm{mod}}$ (Eq.~(\ref{eq:newpot})).
Lines represent the extrapolation of the calculated numbers to the bulk limit.
Experimental values of bulk melting temperature are shown by stars.
%In the case of gold, open symbols show the experimental data on melting of finite-size nanoparticles
%from Ref.~\citenum{Buffat_1975_PRA.13.2287}.
}
\label{figure_Tmelt_NPs}
\end{figure*}

Table~\ref{tab:lattice-const} presents equilibrium lattice constants for silver, gold and titanium calculated
with $U_{\textrm{Gup}}$, $U_{\textrm{Gup}} + U_{\textrm{lin}}$ and $U_{\textrm{Gup}} + U_{\textrm{mod}}$.
The force created by the linear correction causes a uniform strain on the crystals, which become uniformly compressed.
For silver and gold this effect is rather small (the relative change in the lattice parameters is less than 1\%)
while the relative shortening of titanium crystals is about 2.5\%.
This can also be attributed to the very steep linear correction (i.e., the large force) that should
be used to reproduce the experimental bulk melting temperature for Ti.
Note also that geometry optimization of a Ti crystal using the original Gupta potential yields the structure
which is elongated along the [0001] axis %(i.e. along the direction of the lattice parameter $c$)
as compared to the experimental value (the calculated lattice parameter $c = 4.75$~\AA~vs. the experimental value of 4.68~\AA).
Geometry optimization by means of the linear correction results in a uniform compression of the crystal, which
brings $c$ in a better agreement with the experimental value.

The sigmoid-type modification $U_{\textrm{mod}}$ has a small impact on the equilibrium lattice parameters,
which almost coincide with those predicted by the original Gupta potential and agree reasonably well with the experimental results.
Contrary to the linear correction, $U_{\textrm{mod}}$ does not induce strong compression of the Ti crystal
and its lattice parameters obtained by means of $U_{\textrm{mod}}$ are similar to those calculated with $U_{\textrm{Gup}}$.
As discussed above, this is due to the functional form of $U_{\textrm{mod}}$ wherein
the positive contribution of $U_{\textrm{mod}}^{\prime}$ plays a role at small interatomic distances (which span over
a few nearest atomic layers) while the negative contribution of $U_{\textrm{mod}}^{\prime}$ plays a role at larger values of $r$.
%which enhances the interaction with several nearest atomic layers but weakens the interaction with more distant atoms
%-- a large short-range attractive force is compensated by a long-range repulsive force, which together stabilize the crystal.
%thereby stabilizing the crystal.

Figure~\ref{figure_Tmelt_NPs} shows the melting temperatures of finite-size Ag, Au and Ti nanoparticles
as functions of their inverse diameter $D$.
For all the metals, the bulk melting temperature predicted by the original Gupta potential is significantly
lower than the experimental values.
The most illustrative example is titanium (see the right panel of Fig.~\ref{figure_Tmelt_NPs}) whose melting
temperature calculated with $U_{\textrm{Gup}}$ is approximately 1380~K.
It is more than 500~K lower than the experimental value of 1941~K (marked by a star symbol)
which yields the relative discrepancy of about 30\%.
A similar feature has been observed for gold and silver --
the absolute discrepancy is smaller for these metals (about 330 and 100~K, respectively) while the relative
discrepancy for gold is as large as 25\%.
These results justify further the necessity of correcting the Gupta potential to bring the calculated bulk melting
temperatures in closer agreement with the experimental values.
The modification $U_{\textrm{mod}}$ produces a similar effect as the linear correction --
it leads to an increase of melting temperatures of nanoparticles and, as a result, to an increase
of the bulk melting temperatures.
%The experimental data on the melting temperature of nanometer-size systems are very scarce and available
%only for a few metals, including gold.
%In the middle panel of Fig.~\ref{figure_Tmelt_NPs} the results of MD simulations for gold are compared with experimental data on
%the melting temperatures of gold nanoparticles from Ref.~\citenum{Buffat_1975_PRA.13.2287} (open symbols).
%Results obtained with the modified potential are much closer to the experimental values as compared to
%the original Gupta potential.
%However, one may notice that the experimental slope of the $T_{\textrm{m}}(1/D)$ dependence is somewhat different from those
%calculated with both $U_{\textrm{lin}}$ and $U_{\textrm{mod}}$.
%On the other hand, it should be stressed that the experimental melting temperatures of smaller nanoparticles
%are more scattered.
%Moreover, the melting temperatures of the nanoparticles smaller than about 2.5~nm ($1/D > 0.04$~\AA$^{-1}$) start to deviate from
%a linear dependence that follows from the Pawlow law.
%It is a well-known finite-size quantum effect \cite{Gaston_2018_AdvPhysX.3.1401487} related to an increased role
%of geometric structure that may be very different from that of the bulk counterpart.
The new modification leads to an improvement of the calculated bulk melting temperature for the three metals considered.
Good agreement with the experimental values has been obtained for silver and titanium
(the relative discrepancies from the experimental values are 1.5\% and 0.8\%, respectively)
while a somewhat larger discrepancy of about 6\% has been observed for gold.
The reason for this discrepancy is that the sigmoid-type modification increases the slope of
the $T_{\textrm{m}}(1/D)$ dependence for silver and titanium nanoparticles
but it almost does not change the slope for gold nanoparticles.
The utilized parameters of $U_{\textrm{mod}}$ for gold have been chosen such that all the quantities considered 
in this work agree better with experimental data as compared to the original Gupta potential.
A better agreement might be achieved by performing a more detailed analysis of the multi-dimensional parameter surface
of $U_{\textrm{mod}}$.  
A finer tuning of the parameters should bring the calculated $T_{\textrm{m}}^{\textrm{bulk}}$ for gold
to a better agreement with experimental data.

\section{Conclusion and outlook}
\label{sec:conclusions}

We have formulated a recipe for a modification of classical embedded-atom method-type potentials aiming at a
quantitative description of both equilibrium and non-equilibrium properties of metal systems by means of
molecular dynamics simulations.
The modification suggested in this work %has a correct asymptotic behavior
asymptotically approaches zero at large interatomic distances
and generalizes the previously developed linear correction \cite{Sushko_2016_JPCM.28.145201}.
A general procedure for constructing the modified EAM-type potential has been outlined and the relation
between parameters of the new modification and the linear correction was thoroughly elaborated.

The procedure developed has been applied to analyze the melting temperature
%and several equilibrium properties (lattice constants, cohesive energy and vacancy formation energy)
as well as lattice constants, cohesive energy and vacancy formation energy of nanosystems made of silver, gold and titanium.
It was demonstrated that the modified potential leads to an increase of the melting temperature of the metals and
to a better agreement with experimental values as compared to the uncorrected potential.
The Gupta potential has been chosen as a case study but the generality of the correction allows its
application in combination with other widely-used potentials of the EAM type such as Sutton-Chen or Finnis-Sinclair potentials.
The results presented for the metals with cubic and hexagonal crystalline lattices confirm further a wide range
of applicability of the proposed modification.

One of the not yet resolved questions concerns the physical nature of the effects that are produced by the
modified potential.
As it has been demonstrated for the case of titanium, the new modification induces a small (on the order of a few per cent or less)
change of the equilibrium properties but increases the bulk melting temperature by more than 30\%.
This may be attributed to the formation of small nanoclusters which have a different crystalline order
and coordination number compared to the bulk crystal.
Related phenomena were discussed in a recent paper \cite{Nasibullin_2019_CPL.716.199} devoted to investigation
of atomistic-level mechanisms of martensite phase trasitions in NiTi alloys.
It was discussed that such phase transitions are preceded by the formation of pre-martensite states, which are
characterized by the presence of nanometer-size domains with the pentagonal symmetry, characteristic of small icosahedral clusters.
Similar domains may be formed in fcc and hcp metal systems interacting via the modified potential
at elevated temperatures.
This hypothesis alongside with other possible explanations of the effects observed should be thoroughly elaborated,
and we hope this can be addressed in our future work.

%%%%%%%%%%%%%%%%%%%%%%%%%%%%%%%%%%%%%%%%%%%%%%%%%%%%%%%%%%%%%%%%%%%%%
%%%%%%%%%%%%%%%%%%%%%%%%%%%%%%%%%%%%%%%%%%%%%%%%%%%%%%%%%%%%%%%%%%%%%
\appendix*
\section{Derivation of parameters of $U_{\textrm{mod}}$}
\label{sec_appendix}

To analytically derive the parameters of the sigmoid-type potential $U_{\textrm{mod}}(r)$,
the latter was approximated by a piecewise linear function:
\begin{equation}
\bar{U}_{\textrm{mod}}(r) =
\left\{
\begin{array}{l l }
%\begin{gathered}
B_1 \, r + C_1 \ \ , \ r \le R_0 \\
B_2 \, r + C_2 \ \ , \ R_0 < r < R_2 \\
0 \ \ \ \ \qquad \quad , \ r \ge R_2 \\
%\end{gathered} \right.
\end{array} \right.
\label{eq:newpot_approximation}
\end{equation}
where $B_1 > 0$ ($C_1 < 0$) and $B_2 < 0$ ($C_2 > 0$),
$R_0 = - \frac{C_1 - C_2}{B_1 - B_2}$ is the point of intersection of the two linear segments, and
$R_2 = - C_2/B_2$ is the point where $\bar{U}_{\textrm{mod}}(r)$ is equal to zero (see dotted curves in Fig.~\ref{figure_newpot_alphas}).
%
%Similar to Eq.~(\ref{Lin_cond1_general}), the contribution of the sigmoid-type correction to the
%total potential energy of a system should be equal to zero to not change its cohesive energy. %Therefore,
%This condition can be expressed as
%\begin{eqnarray}
%\bar{U}_{\textrm{mod}}^{\textrm{tot}}
%= \int\limits_{r < R_2} n_0 \, \bar{U}_{\textrm{mod}}(r) \, \textrm{d}V %+ \int\limits_{R_0 < r < R_2} n_0 \, \bar{U}_{\textrm{mod}}(r) %\,
%= 0 \ .
%\label{Newpot_cond1_general}
%\end{eqnarray}
%
After substituting (\ref{eq:newpot_approximation}) into Eq.~(\ref{Lin_cond1_general}) and carrying out the integration
%and substituting the expressions for $R_0$ and $R_2$
one arrives at the following condition:
%\begin{equation}
%\frac{(C_1 - C_2)^4}{(B_1 - B_2)^3} = - \frac{C_2^4}{B_2^3} \ ,
%\label{Newpot_cond1_2}
%\end{equation}
%which can be rewritten as
\begin{equation}
\frac{(1 - {\cal C})^4}{(1 - {\cal B})^3} = - \frac{{\cal C}^4}{{\cal B}^3} \ ,
\label{Newpot_cond1_2b}
\end{equation}
where ${\cal B} = B_2/B_1 < 0$ and ${\cal C} = C_2/C_1 < 0$.
Substituting $\bar{U}_{\textrm{mod}}(r)$ in Eq.~(\ref{eq:linear_condition2}) one derives
the force $F_{\textrm{mod}}$ due to this potential.
%
%A condition on the force due to $\bar{U}_{\textrm{mod}}$ acting on an atom located at the origin gives
%\begin{eqnarray}
%F_{\textrm{mod}} = - (B_1 - B_2) \, n_0 \, \frac{2\pi}{3} \, R_0^3 - B_2 \, n_0 \, \frac{2\pi}{3} \, R_2^3
%\end{eqnarray}
%
The change in total potential energy %associated with the displacement of an atom by $\Delta r$ then reads
due to displacement of an atom from its original equilibrium position by $\Delta r$ reads:
\begin{equation}
\Delta U
= - F_{\textrm{mod}} \, \Delta r
= - \frac{2\pi}{3} \, n_0 \, \frac{C_1^3}{B_1^2} \left[ \frac{(1 - {\cal C})^3}{(1 - {\cal B})^2}
+ \frac{{\cal C}^3}{{\cal B}^2} \right] \, \Delta r \ .
\label{Newpot_cond2}
\end{equation}

The force $F_{\textrm{mod}}$ should be equal to the force $F_{\textrm{lin}}$ arising due to the
linear correction at a given cutoff in order to increase the melting temperature by the same value.
Therefore, equating Eq. (\ref{eq:linear_condition2}) to (\ref{Newpot_cond2}) gives
\begin{equation}
\left( \frac{4}{3} \right)^3 \, \frac{C^3}{B^2}
=
\frac{C_1^3}{B_1^2} \left[ \frac{(1 - {\cal C})^3}{(1 - {\cal B})^2}
+ \frac{{\cal C}^3}{{\cal B}^2} \right]  \ .
\label{eq:equal_forces}
\end{equation}
The l.h.s. of this expression depends (according to Eq.~(\ref{Lin_cond1})) on the cutoff distance $r_c$ which
does not have a clear physical meaning but is rather a computational parameter that can be chosen arbitrary.
The r.h.s., on the contrary, depends on the parameters $B_{2}$ and $C_{2}$ (through ${\cal B}$ and ${\cal C}$)
which define the physical range of $\bar{U}_{\textrm{mod}}(r)$ at which the interatomic interactions vanish.
Defining the range of the potential $\bar{U}_{\textrm{mod}}(r)$ is thus equivalent to the choice of cutoff in
the case of the linear correction $U_{\textrm{lin}}(r)$.

%Because of the different slopes of the potentials $\bar{U}_{\textrm{mod}}(r)$ and $U_{\textrm{lin}}(r)$,
%the distances at which they are equal to zero can be different.
%These distances are interrelated as
%$\frac{C_1}{B_1} = \alpha \frac{C}{B}$,
%where $\alpha$ is a proportionality factor.
%After substituting this relation into Eq.~(\ref{eq:equal_forces}) the latter becomes
%the function of only one parameter of the linear correction:
%\begin{equation}
%\left( \frac{4}{3} \right)^3
%=
%\alpha^3 \, \frac{B_1}{B} \left[ \frac{(1 - {\cal C})^3}{(1 - {\cal B})^2}
%+ \frac{{\cal C}^3}{{\cal B}^2} \right]  \ .
%\label{eq:equal_forces2}
%\end{equation}

The procedure for deriving the parameters of the sigmoid-type function $U_{\textrm{mod}}$ (\ref{eq:newpot})
and its approximation $\bar{U}_{\textrm{mod}}$ (\ref{eq:newpot_approximation}) can be summarized as follows. \\
(i) First, the parameters $B$ and $C$ of the linear correction are obtained as described in Section~\ref{sec:linear-correction}. \\
(ii) Then, fixing the point $R_1 = - C_1/B_1$ at which $U_{\textrm{mod}}(r) = 0$ (see Fig.~\ref{figure_newpot_alphas})
%(i.e. fixing the value of $\alpha$)
a scan over different values of $B_1$ and $C_1$ is performed. \\
% and the corresponding values of $C_1$ are obtained. \\
(iii) Next, ${\cal B}$ and ${\cal C}$ are derived from the numerical solution of Eqs. (\ref{Newpot_cond1_2b})
and (\ref{eq:equal_forces}), %\\
%(\ref{eq:equal_forces2}).
%This system of equations has %six
%several pairs of roots (${\cal B},чё {\cal C}$), some of which satisfy the constraints imposed
%on ${\cal B}$ and ${\cal C}$ that both quantities should be real negative numbers.
%These constraints are fulfilled for every $\alpha$ in a rather narrow ``band'' of $B_1$ values (see an illustration in Fig.~\ref{figure_Umod_param}). \\
%(iv) After ${\cal B}$ and ${\cal C}$ have been derived,
and the corresponding values of $B_2$ and $C_2$ are obtained. \\
(iv) Repeating steps (i)-(iii) for different
%values of $\alpha$
combinations ($B_1,C_1$)
one obtains a multi-dimensional surface $(B_1,C_1,B_2,C_2)$. \\
%which satisfies the above-described conditions. \\
(v) Once the parameters $B_{1,2}, C_{1,2}$ are derived, the resulting piece-wise function is fitted with 
the sigmoid-type function $U_{\textrm{mod}}$ (\ref{eq:newpot}) to obtain the parameters $\lambda$ and $r_s$.

\begin{figure*}[htb!]
\centering
\includegraphics[width=0.9\textwidth,clip]{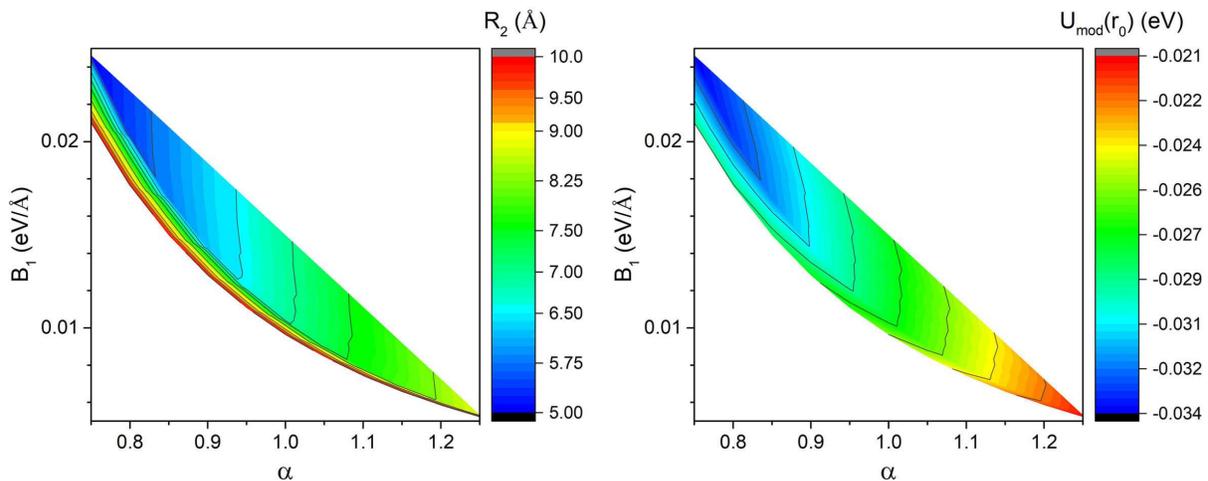}
\caption{
Contour plots for the parameters of $\bar{U}_{\textrm{mod}}(r)$ for silver.
Left panel: a plot of $R_2 = -C_2/B_2$ as a function of $B_1$ and $\alpha$ (see the text for details).
For $\alpha \approx 0.8$ both segments of the piecewise function, Eq.~(\ref{eq:newpot_approximation}), are very steep,
which corresponds to large values of $B_1$ and small values of $R_2$.
For $\alpha \approx 1.2$, the interactions span over larger interatomic distances of about 8~\AA.
Right panel: a plot for $U_{\textrm{mod}}(r_0)$ calculated at the equilibrium point for the original Gupta potential,
as a function of $B_1$ and $\alpha$.
Small values of $\alpha$ yield the largest deviation from the original potential energy curve.
}
\label{figure_Umod_param}
\end{figure*}

The above outlined procedure gives a multi-dimensional parameter surface and additional considerations
should be taken into account to narrow the range of parameters of $U_{\textrm{mod}}$.
The parameters of the Gupta potential considered in this work were derived \cite{Cleri_1993_PRB.48.22}
accounting for interatomic interactions up to the fifth-neighbor shell for fcc metals and up to
seven or eight shells for hcp metals (see Fig.~\ref{figure_B-C_linear} and the discussion in Sect.~\ref{sec:linear-correction}).
The typical range of the sigmoid-type potential should therefore span over five to nine layers of neighboring atoms
and it should smoothly decrease to zero at larger interatomic distances.
This condition imposes a limit on the value $R_2 = -C_2/B_2$ at which $\bar{U}_{\textrm{mod}}(r) = 0$.
The left panel of Fig.~\ref{figure_Umod_param} shows a contour plot of $R_2$ as a function of $B_1$ and the parameter
$\alpha = \frac{C_1}{C} \frac{B}{B_1}$, which defines how steep is $U_{\textrm{mod}}$
(and the first segment of $\bar{U}_{\textrm{mod}}$) with respect to the linear correction $U_{\textrm{lin}}$.
%both of which govern the force acting on the nearest neighboring layers.
The slope of the first segment of $\bar{U}_{\textrm{mod}}$ decreases with an increase of $\alpha$, see Fig.~\ref{figure_newpot_alphas}.
When $\alpha < 1$ (the top-left corner in the left panel of Fig.~\ref{figure_Umod_param}),
both parts of the piecewise function (\ref{eq:newpot_approximation})
are typically very steep, which corresponds to large values of $B_1$ and small values of $R_2$.
%A steep decay of the sigmoid part creates a strong additional force acting on the second and third
%layers of neighboring atoms ($R_2 \approx 5$~\AA), which leads to non-physical effects.
For $\alpha < 1$ one can also derive the parameters of $\bar{U}_{\textrm{mod}}(r)$ such that its range would span up to 10~\AA;
this part of the multi-dimensional surface represents a narrow stripe shown by red color.
%
%For $\alpha > 1$, the forces acting on the nearest neighbors and more distant atoms due to $\bar{U}_{\textrm{mod}}(r)$
%both the attractive and the repulsive forces arising due to $\bar{U}_{\textrm{mod}}(r)$
%are smaller in absolute values.
%In this case,
With an increase of $\alpha$ the range of the sigmoid-type correction increases and the interactions span over
larger interatomic distances.
For instance, for $\alpha = 1.2$ the smallest value of $R_2$ is approx. 8~\AA.

%that makes the computations more CPU-time consuming.
%From these considerations, %more attention has been paid to the region $\alpha \approx 0.95 - 1.10$.
%the region of $\alpha \approx 0.95 - 1.10$ has been chosen for a more detailed investigation.

Another important constraint is that the total potential energy $U$, Eq.~(\ref{eq:newpot_add}),
evaluated at the equilibrium point $r_0$ of the original Gupta potential should change as little as possible
to keep the equilibrium properties close to the values predicted by the original potential.
The right panel of Fig.~\ref{figure_Umod_param} shows a contour plot for $\bar{U}_{\textrm{mod}}(r_0)$
calculated at the equilibrium point for the original Gupta potential.
Parameters corresponding to the small values of $\alpha \approx 0.7-0.8$ yield the largest deviation
from the original potential energy curve, while a smaller impact on the near-equilibrium properties
can be achieved using the parameters that correspond to the values $\alpha > 1$.
% should make a smaller impact on the near-equilibrium properties but their use is also limited by computational efficiency.
%Therefore, the region of intermediate values of $\alpha$ around unity has been chosen for a more detailed investigation.

%%%%%%%%%%%%%%%%%%%%%%%%%%%%%%%%%%%%%%%%%%%%%%%%%%%%%%%%%%%%%%%%%%%%%
\section*{Acknowledgements}

This project has received funding from the European Union's Horizon 2020 research and innovation programme
under grant agreement No 794733 (H2020-MSCA-IF-2017 ``Radio-NP'').
This work was also supported by the Deutsche Forschungsgemeinschaft (Project no. 415716638) and
the Alexander von Humboldt Foundation Linkage Grant.
The possibility to perform computer simulations at the Goethe-HLR cluster of
the Frankfurt Center for Scientific Computing is gratefully acknowledged.

%%%%%%%%%%%%%%%%%%%%%%%%%%%%%%%%%%%%%%%%%%%%%%%%%%%%%%%%%%%%%%%%%%%%%

\end{document}